\documentclass[superscriptaddress,prl,twocolumn,showpacs,amsmath,letter]{revtex4}
\usepackage{graphicx}

\newcommand{\Journal}[4]{#1 \textbf{#2}, #3 (#4)}

\begin{document}

\title{Giant Magnetoresistance in Multilayers with Noncollinear Magnetizations}
\author{S. Urazhdin}
\affiliation{Department of Physics and Astronomy, Johns Hopkins
University, Baltimore, MD, 21218} \affiliation{Department of
Physics and Astronomy and Center for Fundamental Materials
Research, Michigan State University, East Lansing, MI, 48824}
\author{R. Loloee}
\affiliation{Department of Physics and Astronomy and Center for
Fundamental Materials Research, Michigan State University, East
Lansing, MI, 48824}
\author{W.P. Pratt Jr.}
\affiliation{Department of Physics and Astronomy and Center for
Fundamental Materials Research, Michigan State University, East
Lansing, MI, 48824}


\begin{abstract}
We study the dependence of perpendicular-current magnetoresistance in magnetic
multilayers on the angle between the magnetizations of the layers.
This dependence varies with the thickness of one of the layers,
and is different for multilayers with two and three magnetic
layers. We derive a system of equations representing an extension
of the two-current series resistor model, and show that the
angular dependence of magnetoresistance gives information about
the noncollinear spin-transport in ferromagnets.
\end{abstract}

\maketitle

The discoveries of giant magnetoresistance
(GMR)~\cite{gmrdiscovery} and spin-transfer~\cite{tsoiprl} in
ferromagnetic metallic multilayers greatly contributed to our
understanding of the relation between magnetism, charge, and spin
transport, and lead to important applications in
memory devices and sensors. The spin-torque theory of
spin-transfer relies on the absorption of the transverse
spin-current at the magnetic interfaces in multilayers with
noncollinear magnetizations, due to the averaging of
spin-dependent electron reflection at the interfaces, and
spin-precession inside the ferromagnets
~\cite{slonczewski,slonczewski3,stilesapl,kovalev,shpiro,bauer,bauerbook,nonmonotonic,stilesprivate,vedyayev}.
Similarly, the disappearance of the transverse spin-current inside
ferromagnets is predicted to affect the angular dependence of perpendicular-current (CPP) GMR
in multilayers with noncollinear magnetizations (AGMR). Thus, AGMR
is an important effect complimentary to the spin-torque.

Theories of AGMR qualitatively agree with the available
data~\cite{vedyayev,fert96,dieny96,prattunpublished}, but
quantitative agreement has not yet been achieved. Due to the lack
of systematic studies, it is also impossible to verify the
predicted tendencies for the variation of AGMR with the multilayer
parameters. Here, we present a systematic study of AGMR
in multilayers with two and three magnetic layers, in which we
varied the thickness of one of the layers. The
dependence of AGMR on this thickness is different in the two
studied structures. Our analysis shows that
magnetic interfaces, transverse to magnetizations spin-currentsw in ferromagnets, and
in some cases sample leads give contributions to AGMR.
These findings are important for the understanding of
spin-transport in ferromagnets and theories of
spin-torque.

\begin{figure}
\includegraphics[scale=0.45]{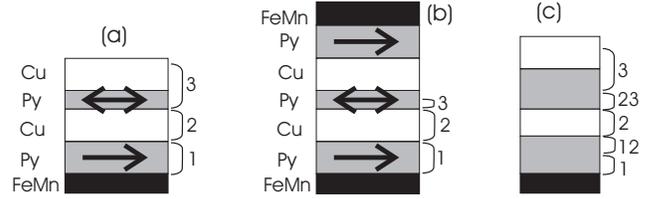}
\caption{\label{fig1} (a,b) Schematics of samples A (a) and B (b),
with layer compositions as labeled (thicknesses not to scale).
Different sections of samples are labeled 1-3, as described in the
text. (c) Schematic of the model incorporating finite transverse
spin-current decay length.}
\end{figure}

Our sample fabrication and measurement techniques were described
elsewhere~\cite{technique}. The structure of sample type A was
Nb(150)Cu(20)FeMn(8)Py(6)Cu(10)Py(1.5-12)Cu(20)Nb(150),
Py=Permalloy=Ni$_{84}$Fe$_{16}$ (Fig.~\ref{fig1}). We label these
samples A1.5, A3, A6, A12 by the thickness $t_{Py}$ of the top Py
layer. All thicknesses are in nanometers. Samples B1.5-12 had
structure
Nb(150)Cu(20)FeMn(5-12)Py(6)Cu(10)Py(1.5-12)Cu(10)Py(6)FeMn(5-12)Cu(20)Nb(150),
and were labeled by the thickness of the middle Py layer. The
bottom Py(6) in samples A, and the outer Py(6) layers in samples B
were exchange-biased by annealing in magnetic field $H\approx
30$~Oe at $170^\circ$C. At least 2 samples of each type were
measured with similar results.

\begin{figure}
\includegraphics[scale=0.75]{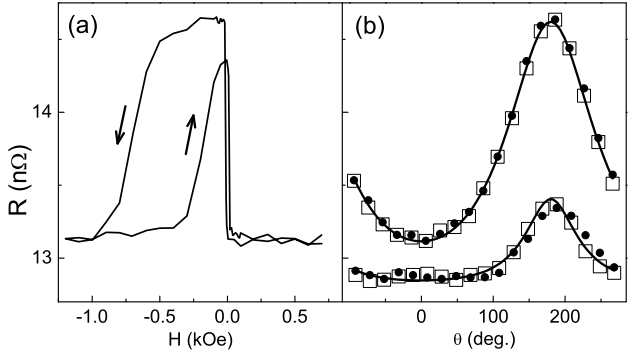}
\caption{\label{fig2} (a) $R$ {\it vs.} $H$ in sample A12. (b)
AGMR at $H=25$~Oe (dots) and $H=100$~Oe (squares) in samples A1.5
(bottom) and A12 (top). Solid lines are fits of the $100$~Oe data
with $\chi =7.7\pm 0.6$ and $\chi =1.96\pm 0.05$, respectively, as
described in the text. For clarity, not all the measured points
are shown, and the A1.5 data are offset by $-0.2n\Omega$.}
\end{figure}

Sample resistances $R$ were measured with a SQUID voltmeter at
4.2~K in a CPP geometry. Fig.~\ref{fig2}(a) shows an example of $R$ {\it vs.} $H$ curve for
sample A12, measured with $H>0$ along the pinning direction
of the bottom Py layer. At large $H>0$, magnetizations of both
Py layers are parallel (P) to each other and $H$, giving low
resistance $R_P$. As $H$ decreased to small $H<0$, the free layer
switched to give antiparallel (AP) state with high resistance
$R_{AP}$. The free layer coercive field $H_c$ (half the width of
its hysteresis) decreased from $\approx 30$~Oe in samples A1.5 to
$10$~Oe in A12. The decrease of resistance to $R_P$ at
significantly larger $H<0$ is due to switching of the pinned Py
layer. We define the bias field $H_b$ as $H$ necessary to achieve
$R=(R_P+R_{AP})/2$. In most samples, $|H_b|\approx 800~Oe$.
Because $H_c\ll H_b$, we were able to rotate the free Py layer's
magnetization, not significantly affecting the pinned one.

Fig.~\ref{fig2}(b) shows examples of the AGMR measurements for
samples A1.5 and A12, performed by rotating a fixed $H=25-100$~Oe
in the plane of the films. At angle $\theta=0$, $H$ is in the
pinning direction of the bottom Py. There was no significant
dependence of data on $25\le H\le 100$~Oe in samples A3-A12. In
samples A1.5, $H=25$~Oe was insufficient to completely reorient
the Py(1.5) layer due to its higher coercivity, but at $50\le H\le
100$~Oe data were independent of $H$. Thus, we conclude that: a)
our $H\le 100$~Oe~$\ll H_b$ does not significantly affect the
magnetization of the pinned layer, b) Except for samples A1.5,
$H=25$~Oe was sufficient for monodomain rotation of the free
layer.

\begin{figure}
\includegraphics[scale=0.41]{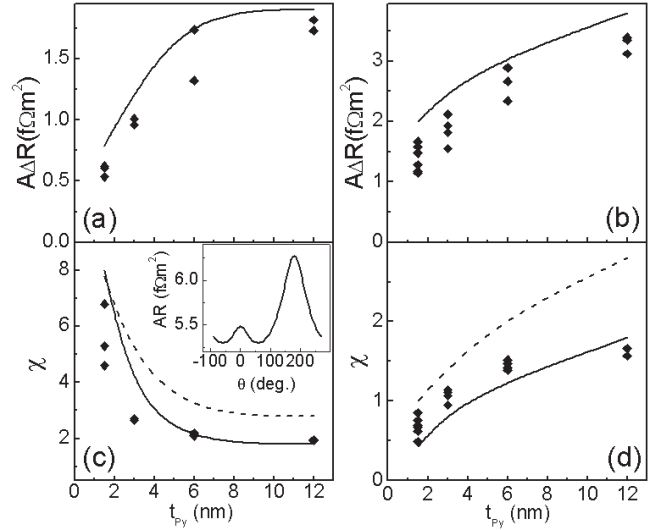}
\caption{\label{fig3} (a) Measured (symbols) and calculated (line)
$A\Delta R$ {\it vs.} $t_{Py}$ for samples A. (b) Same as (a), for
samples B. (c) $\chi$ from the fits of measured (symbols) and
calculated (lines) AGMR with Eq.~\ref{chifit} for samples A. Solid
line is calculated with correction for a finite $l_t=0.8$~nm,
dashed line -- with $l_t=0$. Inset: calculated $R(\theta)$ for
A1.5, as described in the text. (d) Same as (c), for samples B. All
the lines are B-spline fits of the calculated points for
$t_{Py}=1.5,3,6,12$~nm.}
\end{figure}

Solid lines in Fig.~\ref{fig2}(b) are fits of the $100$~Oe data
with
\begin{equation}\label{chifit}
R(\theta)=R_P+\Delta R\frac{1-\cos^2(\theta/2)}{1+\chi
\cos^2(\theta/2)},
\end{equation}
proposed by Giacomoni {\it et al.}~\cite{prattunpublished}, and
later derived for symmetric
spin-valves~\cite{slonczewski3,stilesapl,kovalev,shpiro,bauer,bauerbook,nonmonotonic,stilesprivate}.
In Eq.~\ref{chifit}, $\Delta R=R_{AP}-R_P$, and $\chi$ is a
fitting parameter. Eq.~\ref{chifit} gives good fit for all samples
except for A1.5. In all three samples A1.5, the best fit did not
completely reproduce the nearly constant data for
$-90^\circ<\theta<90^\circ$. Because of the finite $\theta=0$
curvature of the fit, it was below the data at $\theta\approx 0$,
and slightly above it around $\theta=100^\circ$. Data for A1.5 in
Fig.~\ref{fig3}(b) show a weak rise around $\theta=0$, similar to
the uncertainty of the measurements.

Both $R(H)$ and AGMR data for samples B were qualitatively similar
to those in Fig.~\ref{fig2}. Figs.~\ref{fig3}(a,b) summarize the
$A\Delta R$ measurements (where $A$ is the sample area) for samples
A and B, and Figs.~\ref{fig3}(c,d) show the $\chi$ values, extracted
by fitting the AGMR data with
Eq.~\ref{chifit}. To reduce the uncertainty, $\chi$ obtained from
$100$~Oe and $50$~Oe data were averaged for each point. At large
$t_{Py}$, sample B is expected to behave similarly to two samples
A connected in-series. Consistently, Fig.~\ref{fig3} shows that
$A\Delta R$ for samples B12 is double that for samples A12, while
$\chi$s are close. At smaller $t_{Py}$, $A\Delta R$ in samples A
decreases faster than half of that in samples B. In samples
A, $\chi$ increases, while in samples B it decreases at smaller $t_{Py}$.

Several models have treated
AGMR~\cite{slonczewski3,stilesapl,kovalev,shpiro,bauer,bauerbook,nonmonotonic,stilesprivate,vedyayev}.
They predict the form Eq.~\ref{chifit} for symmetric spin-valves,
but overestimate $\chi$ typically by a factor of two. Some
theories predict a negligible length scale $l_t$ for the decay of
the transverse spin-current in Py~\cite{slonczewski3,stilesapl}.
Others correlate $l_t$ with the magnetic length $\approx4$~nm in
Py~\cite{shpiro}.

We develop an extension of the two current series resistor
model~\cite{tworesistor} (2CSRM) to AGMR, which contains much of
the same physics as these other models, but allows us to gain
qualitative insight into our results, analyze the effects of
finite $l_t$, and include the effect of the Nb sample leads, which
we shall see give an important contribution to the measured
dependencies.

In 2CSRM, one separately considers two spin-current channels, same
across the whole sample. For noncollinear magnetizations, two
common spin-channels across the multilayer generally do not exist,
so one needs to consider how the spin-channels are transformed
across the multilayer. In samples A, we consider three parts of
the multilayer (labeled 1-3 in Fig.~\ref{fig1}(a)), separated by
the Py/Cu interfaces 12 and 23. The outer limits of regions 1 and
3 are determined by the spin-diffusion, i.e. the GMR-active part
of the multilayer. This additional constraint augments 2CSRM and
is essential for the following analysis. At each point of the
multilayer, we define a matrix current $\hat{I}=\left[
\begin{array}{rr} I^{11} & I^{12}\\ I^{21} & I^{22}\\
\end{array}\right]$, which gives charge current $I_c=Tr \hat{I}$ and projection of spin
current on an arbitrary axis $\mathbf{m}$, $I_s(\mathbf{m})=\hbar
Tr(\sum_i m_i\sigma_i\hat{I})/(2e)$. Matrix product of $\hat{I}$ and the Pauli
matrices $\sigma_i$  is implied here. We
assume that, due to averaging of spin-precession in the
ferromagnets, the component of
spin-current transverse to the Py magnetizations vanishes in regions 1 and
3~\cite{slonczewski,slonczewski3,stilesapl}. Later, we will
discuss and modify this assumption. In region 1, the matrix
current $\hat{I}_1$ is then diagonal in the frame set by the
magnetization $M_1$ of the bottom Py layer, $\hat{I_1}=diag
\hat{I}_1$. Similarly, in region 3, $\hat{I}'_3=diag \hat{I}'_3$.
We use primed symbols for the frame set by the orientation of the
top Py magnetization $M_2$.

If spin-flip scattering within regions 1-3 is neglected, $\hat{I}$
is conserved separately across each region. However, only $I_c$
and the spin-current projections on the corresponding Py
magnetizations are conserved at the interfaces 12 and 23,
therefore $\hat{I}_1=diag \hat{I}_2$, $\hat{I'}_3=diag
\hat{I'}_2$. Here $\hat{I'}=\hat{U}^+\hat{I}\hat{U}$,
$\hat{U}=\left[
\begin{array}{rr} \cos(\theta/2) & -\sin(\theta/2)\\ \sin(\theta/2) & \cos(\theta/2)\\
\end{array}\right]$ is the spin rotation matrix by angle $\theta$ between $M_1$ and $M_2$.

We describe the local electron distributions by $2\times 2$ spinor
distribution matrices~\cite{bauer,bauerbook}. Their diagonal
elements in an arbitrary reference frame are given by the spin-up
and spin-down electron densities, related to this frame. We
neglect scattering in the Cu spacer between the Py layers, i.e.
assume a position independent distribution in region 2. Finally,
we introduce matrix resistances
$\hat{R_1}=2R_1^*\left[ \begin{array}{rr} 1-\beta_1 & 0\\ 0 & 1+\beta_1\\
\end{array}\right]$, $\hat{R}_{3}=
2R_3^*\hat{U}\left[\begin{array}{rr} 1-\beta_3 & 0\\ 0 & 1+\beta_3\\
\end{array}\right]\hat{U}^+$, $\hat{R}_{12}=2R^*_{12}\left[ \begin{array}{rr} 1-\gamma & 0\\
0 & 1+\gamma\\ \end{array}\right]$, and
$\hat{R}_{23}=\hat{U}\hat{R}_{12}\hat{U}^+$ (for identical
interfaces 12 and 23), which connect martrix currents across
regions 1 and 3, and interfaces 12 and 23, to the corresponding
variations of the electron distributions. $R^*$, $\gamma$, and $\beta$
are standard GMR notations~\cite{tworesistor}.

The total voltage across the multilayer $\hat{V}=\left[
\begin{array}{rr} V & 0\\ 0 & V\\
\end{array}\right]$ is the sum of electron distribution variations
across regions 1 and 3, and interfaces 12 and 23,
\begin{eqnarray}\label{matrixOhms}
\nonumber\hat{V}=\hat{R_1}\hat{I}_1+(\hat{R}_{12}+\hat{R}_{23})
\hat{I}_2+\hat{R_3}\hat{I}_3=
\hat{R_1}diag(\hat{I}_2)+\\(\hat{R}_{12}+\hat{R}_{23})
\hat{I}_2+\hat{R_3}\hat{U}diag(\hat{U}^+\hat{I}_2\hat{U})\hat{U}^+.
\end{eqnarray}
Eq.~\ref{matrixOhms} connects four unknown components of
$\hat{I}_2$ to the voltage $\hat{V}$ across the mutlilayer. The
diagonal components of Eq.~\ref{matrixOhms} reduce to 2CSRM in the
collinear limit. Once Eq.~\ref{matrixOhms} is solved for
$\hat{I_2}$, the resistance of the multilayer is given by
$R(\theta)=V/Tr\hat{I}_2$. We note that while Eq.~\ref{matrixOhms}
is written in the frame of $M_1$, it can be transformed to an
arbitrary reference frame.

For a symmetric bilayer, $R_1^*=R_3^*$, $\beta_1=\beta_3$,
Eq.~\ref{matrixOhms} is diagonal in the frame rotated halfway
between $M_1$ and $M_2$. In this special case
\begin{equation}\label{Rsymmetric}
R(\theta)=2R_1^*+2R_{12}^*-\frac{2\cos^2(\theta/2)(R_1^*\beta_1+R_{12}^*\gamma)^2}
{R_{12}^*+R_1^*\cos^2(\theta/2)}.
\end{equation}
This expression has the same form as Eq.~\ref{chifit}, with
$\chi=R_1^*/R_{12}^*$. As we noted above, $R_1^*$ and $\beta_1$
are determined by the condition that only the GMR-active part of
the multilayer is included. In particular, in samples A12
$t_{Py}>l_{sf}\approx 5.5$~nm, the spin-diffusion length in
Py~\cite{mrparameters}, so only the $l_{sf}$-thick parts of the Py
layers should be included in regions 1 and 3. Thus, samples A12
are approximately symmetric, and can be described by
Eq.~\ref{Rsymmetric}. Similar arguments hold for samples B12,
viewed as two samples A connected in series. We use the parameters
established in CPP-GMR measurements~\cite{mrparameters},
$AR_1^*=l_{sf}\rho^*_{Py}=1.4$~f$\Omega$m$^2$,
$\beta_1=\beta_{Py}=0.7$, $AR_{12}^*=0.5$~f$\Omega$m$^2$,
$\gamma=0.7$. We obtain $\chi=R_1^*/R_{12}^*=2.8$,
larger than the measured $\chi = 2.0$ for A12, and $\chi = 1.6$
for B12.

Overestimation of $\chi$ is a general tendency of the AGMR
theories. The physical transparency of Eq.~\ref{Rsymmetric} allows
us to identify the possible origins of this discrepancy, and
appropriately correct our analysis. We assumed above that
transverse spin-current and the nondiagonal electron distribution
components vanish in Py arbitrarily close to the Py/Cu interface.
However, a sharp Py/Cu interface may be an unjustified
idealization. If the onset of the bulk ferromagnetic Py properties
occurs over a finite thickness, where Py and Cu are alloyed, it is
also reasonable to expect that the transverse spin-current decays
over a finite length $l_t$, nominally inside Py. Moreover, some
theories predict a finite $l_t$ even when an ideally sharp
interface is assumed~\cite{shpiro}. Regardless of the physical
origin, we can phenomenologically include a finite $l_t$ into our
model by expanding the interfaces 12 and 23 into finite
$l_t$-thick regions of Py (Fig.~\ref{fig1}(c)), where spin-current
is noncollinear to the magnetization. This correction decreases
$R^*_1$, and increases $R^*_{12}$, thus decreasing $\chi$.
$l_t\approx 0.8$~nm gives a good agreement of calculated $\chi$
with the data for samples A12 and B12.

$A\Delta R$ is not affected by the finite $l_t$, irrelevant for
the collinear transport. For $t_{Py}$ comparable to $l_t$, the
spin-torque should decrease due to incomplete transverse
spin-transfer between electrons and magnetization. We note that in
the published studies of spin-transfer (mostly with Co) the
ferromagnet thicknesses were larger than
$l_t$~\cite{cornellquant}. $l_t$ in Co is likely even smaller than
in Py due to its larger exchange splitting~\cite{shpiro}. The
circuit theory of spin-polarized transport and
spin-transfer~\cite{kovalev,bauer,bauerbook,nonmonotonic} uses a
mixing conductance parameter $g_{\uparrow\downarrow}$,
characterizing spin-dependent scattering at the interfaces.
Related to our model, $2g_{\uparrow\downarrow}=(R_{12}^*)^{-1}$,
thus it depends on $l_t$.

Giacomoni {\it et al.}~\cite{prattunpublished} obtained
$\chi\approx 1.2$ for thick Py/Cu/Py spin-valves similar to our
samples A12 or B12. Their smaller $\chi$ may be due to stronger
alloying of the Py/Cu interfaces (giving larger $l_t$), caused by
higher deposition rates and longer annealing time during pinning.

We modeled the $\chi$ dependence on $t_{Py}$ by solving the
general form of Eq.~\ref{matrixOhms}. The solution for asymmetric
multilayers is similar to Eq.~\ref{chifit}, but more complicated,
and the denominator contains additional $\cos^4(\theta/2)$ terms.
We therefore give only numerical results for specific cases. One has $t_{Py}<l_{sf}$ in
samples A1.5 and A3, so the GMR-active top part of
the multilayer (region 3) must now include the entire top Py layer
and the Py/Cu and Cu/Nb interfaces. Studies of Nb/Py and Nb/Cu/Py
multilayers yield a large interface resistance $AR_{Py/Nb}\approx
3$~f$\Omega$m$^2$~\cite{mrparameters}, both with and without a Cu spacer
between Py and Nb.

For samples A1.5, we added the full value of $R_{Py/Nb}$ to
$\hat{R}_3$, neglecting electron spin-flipping in Py(1.5). For A3,
we added $0.5R_{Py/Nb}$ to approximately account for spin-flipping
in Py(3), reducing the contribution of Py/Nb interface to GMR. Our
model gives the same results for samples A12 and A6, but the
difference in data is also small. The calculated $A\Delta R$
(Fig.~\ref{fig3}(a), solid line) agrees well with the data. For
samples A1.5, the calculated $R(\theta)$ significantly deviates
from the form Eq.~\ref{chifit}; Inset in Fig.~\ref{fig3}(c) shows
that it has maxima both at $\theta=0$ and $\theta=180^\circ$. A
similar behavior is predicted for asymmetric spin-valves by the
circuit theory~\cite{nonmonotonic}. Our calculation exaggerates a
weak rise of data at $\theta\approx 0$ (Fig.~\ref{fig2}(b)), but
captures the overall experimental behavior. The quantitative
discrepancy may be due to the neglected electron spin-flipping at
the interfaces. In Fig.~\ref{fig3}(c), solid line shows the
calculated $\chi$, defined as the best fit with Eq.~\ref{chifit}
to the calculated $R(\theta)$, using $l_t=0.8$~nm. The calculation
using $l_t=0$ (dashed line) gives a significantly worse agreement
with data.

Samples B are symmetric with respect to the center of the middle
Py layer. Therefore, current reversal does not change the electron
distribution at that point for any magnetic orientation of the
middle Py layer. Since the properly offset electron distribution
is proportional to the current, we conclude that in the center of
the middle Py layer the electron distribution is spin-independent.
The model developed above for two magnetic layers can now be
adopted to samples B with $t_{Py}<2l_{sf}$, if we take half of the
middle Py layer as region 3, as shown in Fig.~\ref{fig1}(b). The
top half of the sample simply doubles the resistance
obtained from Eq.~\ref{matrixOhms}, not affecting $\chi$.
The results for $A\Delta R(t_{Py})$ and $\chi(t_{Py})$, with
$l_t=0.8$~nm, are shown with solid lines in Figs.~\ref{fig3}(b,d).
The deviations from the form Eq.~\ref{chifit} were negligible for
all samples B1.5-B12. Our model overestimates $A\Delta R$, but
gives reasonable results for $\chi(t_{Py})$. Calculation assuming
$l_t=0$ (dashed line in Fig.~\ref{fig3}(c)) gives a
worse agreement with data.

Qualitatively, our results for $\chi(t_{Py})$ in both samples A
and B can be understood with Eq.~\ref{Rsymmetric}, derived for
symmetric multilayers. In samples A, the activation of the highly
resistive Py/Nb interface at smaller $t_{Py}$ is equivalent to
an increase of $R^*_1$ in Eq.~\ref{Rsymmetric}, giving larger
$\chi$. In samples B, smaller $t_{Py}$ is
equivalent to reduced $R_1^*$ in the symmetric case, and thus
smaller $\chi$.

In summary, we showed that the variation of GMR with angle between
the magnetic layers (AGMR) depends on the thickness of one of the
magnetic layers. The dependence is different in samples with two
and three magnetic layers. To analyze the data, we developed an
extension of the two current resistor model to multilayers with
noncollinear magnetizations. Our analysis leads to the following
conclusions: i) The deviation of AGMR from sinusoidal behavior is
approximately given by the ratio of two quantities: a) the
resistance of the GMR-active part of the multilayer excluding the
noncollinear ferromagnetic interfaces, b) the resistance of these
interfaces. The magnitude of this effect is not directly related
to magnetic anisotropies and $A\Delta R$; ii) AGMR can be
nonmonotonic in asymmetric spin-valves; iii) the transverse spin
current penetration length $t_l$ into ferromagnet can be extracted
from AGMR. From our model, $l_t\approx 0.8$~nm for Py. $l_t$ is an
important parameter for the models of noncollinear spin transport
in ferromagnets and spin-torque. The spin-torque should be reduced
if the ferromagnet thickness is close to $l_t$.

We acknowledge helpful communications with M.D. Stiles, J. Bass,
N.O. Birge, G.E.W. Bauer, support from the MSU CFMR,
CSM, the MSU Keck Microfabrication facility, the NSF through
Grants DMR 02-02476, 98-09688, and NSF-EU 00-98803.

\end{document}